\documentclass[twocolumn,twoside,slac_two]{revtex4}
\usepackage{graphicx}
\usepackage{fancyhdr}
\newcommand{\Dzro}{D\O\ }

\begin{document}

\title{Properties of Weakly-decaying Bottom Baryons, $\Xi_b^-$ and
       $\Omega_b^-$, at CDF}

%

\author{S. Behari (For the CDF Collaboration)}
\affiliation{The Johns Hopkins University, Baltimore, MD 21218, USA}

\begin{abstract}
We present properties of weakly decaying bottom baryons, $\Xi^-_b$
and $\Omega^-_b$, using 4.2 fb$^{-1}$ of data from $p\bar p$ 
collisions at $\sqrt{s}=1.96$ TeV, and recorded with the Collider 
Detector at Fermilab. We report the observation of the $\Omega^-_b$ 
through the decay chain $\Omega^-_b \rightarrow J/\psi \, \Omega^-$, 
where $J/\psi \rightarrow \mu^+\, \mu^-$, $\Omega^- \rightarrow 
\Lambda \, K^-$, and $\Lambda \rightarrow p \, \pi^-$. Significance
of the observed signal is estimated to be 5.5 Gaussian standard 
deviations. The $\Omega^-_b$ mass and lifetime are measured to be 
$6054.4\pm 6.8 (\textrm{stat.})\pm 0.9 (\textrm{syst.})$ MeV/$c^2$
and $1.13^{+0.53}_{-0.40}(\textrm{stat.})\pm0.02(\textrm{syst.})$ ps,
respectively. In addition, the mass and lifetime of the $\Xi^-_b$ 
baryon are measured to be 
$5790.9\pm2.6(\textrm{stat.})\pm0.8(\textrm{syst.})$ MeV/$c^2$
and $1.56^{+0.27}_{-0.25}(\textrm{stat.})\pm0.02(\textrm{syst.})$ ps,
respectively.
Under the assumption that the $\Xi_b^-$ and $\Omega_b^-$ are produced 
with similar kinematic distributions as the $\Lambda^0_b$ baryon, we 
measure 
$\frac{\sigma(\Xi_b^-){\cal B}(\Xi_b^- \rightarrow J/\psi \, \Xi^-)}
      {\sigma(\Lambda^0_b){\cal B}(\Lambda^0_b \rightarrow J/\psi \, \Lambda)} =
0.167^{+0.037}_{-0.025}(\textrm{stat.})\pm0.012(\textrm{syst.})$ and
$\frac{\sigma(\Omega_b^-){\cal B}(\Omega_b^- \rightarrow J/\psi \, \Omega^-)}
      {\sigma(\Lambda^0_b){\cal B}(\Lambda^0_b \rightarrow J/\psi \, \Lambda)} =
0.045^{+0.017}_{-0.012}(\textrm{stat.})\pm0.004(\textrm{syst.})$
for baryons produced with transverse momentum in the range of
 $6 \, - \,  20$ GeV/$c$.
\end{abstract}

\maketitle

\thispagestyle{fancy}


\section{Introduction}
Until recently, tests of quark model predictions of $b$-baryon 
spectroscopy have been limited to only $\Lambda^0_b$~\cite{PDG}. With 
accumulation of large data sets from the Tevatron, some of the other 
predicted baryons, $\Xi_b^-$~\cite{D0_Xi_b,CDF_Xi_b} and 
$\Sigma_b^{(*)}$~\cite{CDF_Sigma_b} have been observed and are in 
good agreement with the quark model predictions.

In this paper, we report the observation of an additional heavy 
baryon, the doubly-strange $\Omega_b^-$ ($|ssb \, \rangle$),
and the measurement of its mass, lifetime, and relative production
rate compared to the $\Lambda^0_b$ production. Observation of this 
baryon has been previously reported by the \Dzro 
Collaboration~\cite{D0_Omega_b}. However, the analysis presented 
here measures a mass of the $\Omega_b^-$ significantly lower 
than Ref.~\cite{D0_Omega_b}.

The measurements reported here are made using $p\overline{p}$ 
collisions at a center of mass energy of 1.96 TeV acquired by the
Collider Detector at Fermilab (CDF II) and based on a data sample 
corresponding to an integrated luminosity of 4.2 fb$^{-1}$. The
$\Omega^-_{b}$ candidates are reconstructed through the decay chain
$\Omega^-_{b} \rightarrow$ $J/\psi \, \Omega^-$, where
$J/\psi \rightarrow \mu^+\mu^-$, $\Omega^- \rightarrow \Lambda \, 
K^-$, and $\Lambda \rightarrow p \, \pi^-$. Charge conjugate modes
are implied throughout this paper. Mass, lifetime, and production 
rate measurements are also reported for the $\Xi_b^-$, through the
similar decay chain $\Xi^-_{b} \rightarrow$ $J/\psi \, \Xi^-$, where
$J/\psi \rightarrow \mu^+\mu^-$, $\Xi^- \rightarrow \Lambda \, \pi^-$,
and $\Lambda \rightarrow p \, \pi^-$.
The production rates of both the $\Xi^-_b$ and $\Omega_b^-$ are 
measured with respect to the $\Lambda^0_b$, which is observed through 
the decay chain $\Lambda^0_{b} \rightarrow$ $J/\psi \, \Lambda$, where
$J/\psi \rightarrow \mu^+\mu^-$, and $\Lambda \rightarrow p \, \pi^-$.

To build confidence in the analysis procedure, all the measurements 
made here are also performed on better known $b-$hadron states
$B^0 \rightarrow J/\psi \, K^{*}(892)^{0}$, 
$K^{*}(892)^{0}\rightarrow K^+\pi^-$; 
$B^0 \rightarrow J/\psi \, K^{0}_s$,
$K^{0}_s \rightarrow \pi^+ \, \pi^-$; and
$\Lambda^0_b \rightarrow J/\psi \, \Lambda$,
$\Lambda \rightarrow p \, \pi^-$ for comparison with other 
experiments. The $K^{*}(892)^{0}$ mode provides a large $B^0$ sample.
The $K^{0}_s$ is reconstructed  from tracks that are significantly 
displaced from the collision, similar to the final state tracks of 
the $\Xi^-_b$ and $\Omega^-_b$. The $\Lambda^0_b$, on the other hand,
is a suitable reference state for relative production rate 
measurements, since it is the largest sample of reconstructed 
$b$-baryons.

\section{Event Reconstruction\label{sect:Recon}}

\begin{figure}[h]
\centering
\includegraphics[width=80mm]{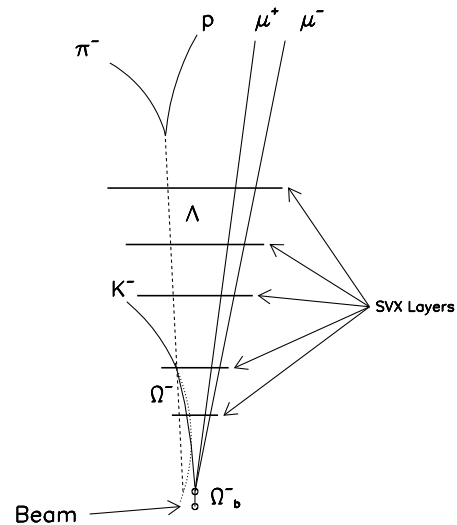}
\caption{An illustration (not to scale) of the $\Omega^-_b \rightarrow 
J/\psi \, \Omega^-$, $J/\psi \rightarrow \mu^+\, \mu^-$, $\Omega^- 
\rightarrow \Lambda \, K^-$, and $\Lambda \rightarrow p \, \pi^-$ final 
state as seen in the view transverse to the beam direction.}
\label{fig:cartoon}
\end{figure}

We employ multi-stage kinematic fits of final state charged particle 
trajectories to infer intermediate and ultimate parent hadron decay
vertices. Fig.~\ref{fig:cartoon} depicts this complex procedure for
the $\Omega^-_b$.
The events are recorded using the CDF di-muon trigger which
is dedicated to collecting $J/\psi \rightarrow \mu^+ \mu^-$ samples.

The event reconstruction begins with a selection of well-measured
$J/\psi \rightarrow \mu^+ \mu^-$ candidates.
This data sample provides approximately $2.9\times 10^7$  $J/\psi$
candidates, measured with an average mass resolution of $\sim20$ MeV/$c^2$.

The $K^0_s$, $K^{*}(892)^{0}$, and $\Lambda$ candidates are reconstructed
from all tracks with $p_T \, >$ 0.4 GeV/$c$ found in the CDF central outer
tracker (COT), that are not associated with muons in the $J/\psi$
reconstruction. Candidate selection for these neutral states is based upon 
the mass calculated for each oppositely charged track pair, which is 
required to fall within $\pm30$, $\pm20$, and $\pm9$ MeV/$c^2$ of the 
nominal mass for the $K^{*}(892)^{0}$, $K^0_s$, and $\Lambda$, respectively.
Backgrounds to the $K^0_s$  and $\Lambda$  are reduced by requiring the 
flight distance of the $K^0_s$ and $\Lambda$ with respect to the primary 
vertex to be greater than 1.0 cm. Approximately $3.6\times 10^6$ $\Lambda$ 
candidates are found with with $p_T(\Lambda) \, > \, 2.0$ GeV/$c$.

\begin{figure}[h]
\centering
\includegraphics[width=80mm]{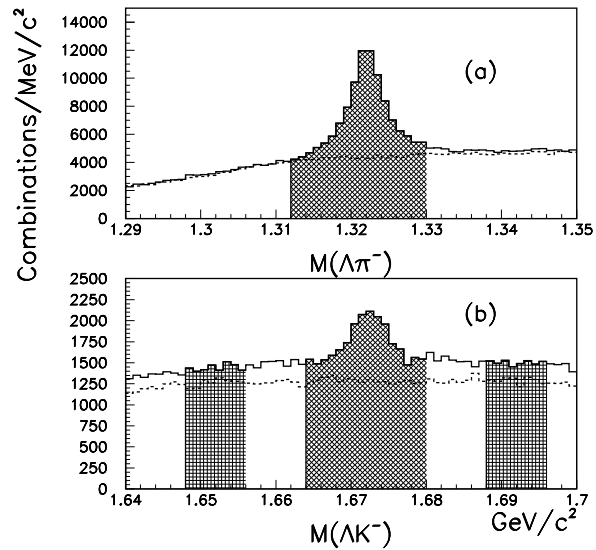}
\caption{The invariant mass distributions of (a) $\Lambda \, \pi^-$ 
combinations and (b) $\Lambda \, K^-$ combinations in events containing 
$J/\psi$ candidates. Shaded areas indicate the signal and sideband 
regions used for $\Xi^-$ and $\Omega^-$ candidates. Wrong-sign
combinations are also shown as dashed histograms.}
\label{fig:incl_hyp}
\end{figure}

Events containing a $\Lambda$ candidate are searched for $\Lambda \, \pi^-$ 
or $\Lambda \, K^-$ combinations consistent with the decay process
$\Xi^- \rightarrow \Lambda \, \pi^-$ or $\Omega^- \rightarrow \Lambda \, 
K^-$ by assigning pion or kaon mass to the remaining tracks. A $p_T(K^-) 
> 1.0$ GeV/$c$ requirement is imposed for our $\Omega^-$ sample, which 
reduces the combinatorial background by 60\%, while reducing the 
$\Omega^-$ signal predicted by our Monte Carlo simulation by 25\%. In 
addition, the flight distance of the $\Lambda$ candidates with respect to 
the reconstructed decay vertex of the $\Xi^-(\Omega^-)$, and the flight
distance from the primary vertex of the $\Xi^-$ and $\Omega^-$ candidates
is required to exceed 1.0 cm. Kinematic reflections are removed from the 
$\Omega^-$ sample by requiring that the combinations consistent with
$\Xi^-$ decay, when the candidate $K^-$ track is assigned the mass of the 
$\pi^-$. Any ambiguities for the proper track assignments of the hadrons 
are resolved examining the $P(\chi^2)$ of the vertex fits. Shown in 
Fig.~\ref{fig:incl_hyp} are approximately $41 \, 000$ $\Xi^-$ and 
$3500$ $\Omega^-$ candidates found in this data sample.
$\Lambda \, \pi^-$ or $\Lambda \, K^-$  combinations within $\pm9$ and 
$\pm8$ MeV/$c^2$ of the nominal $\Xi^-$ and $\Omega^-$ masses are selected 
for $b$-hadron reconstruction. Shown also are the signal and sideband 
regions (shaded) and the wrong-sign combinations (dashed histograms).

Finally $b$-hadron candidates are reconstructed by combining the $K$ and 
hyperon candidates with the $J/\psi$ candidates which involves fitting
the full four-track or five-track state with constraints appropriate
for each decay topology and intermediate hadron state. Specifically, the 
muon pair mass is constrained to the nominal $J/\psi$ mass~\cite{PDG}
and the neutral $K$ or hyperon candidate is constrained to originate 
from the $J/\psi$ decay vertex. In addition, the fits that include the
charged hyperons constrain the $\Lambda$ candidate tracks to the nominal
$\Lambda$ mass~\cite{PDG}, and the $\Xi^-$ and $\Omega^-$ candidates to 
their respective nominal masses~\cite{PDG}. The $\Xi^-_{b}$ and 
$\Omega_b^-$ mass resolutions obtained from simulated events are found to 
be approximately $12$ MeV/$c^2$, a value that is comparable to the mass 
resolution obtained with the CDF II detector for other $b$-hadrons with 
a $J/\psi$ meson in the final state~\cite{CDF_Bmass}.

\section{Observation of the Decay
$\Omega^-_b\rightarrow$ $J/\psi \, \Omega^-$
 \label{sect:signif}}

\begin{figure}[h]
\centering
\includegraphics[width=80mm]{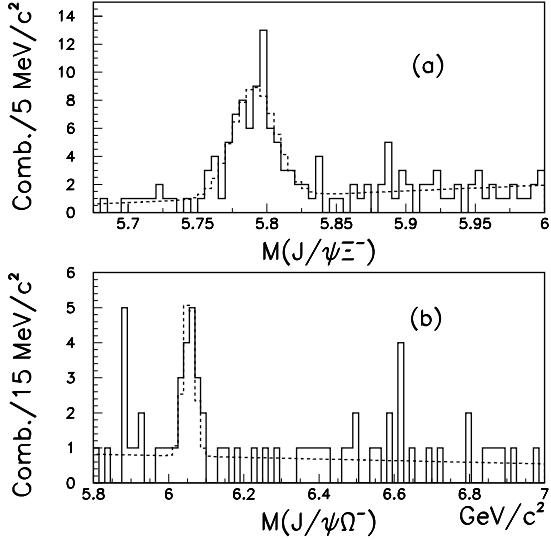}
\caption{The invariant mass of (a) $J/\psi \, \Xi^-$  and (b) $J/\psi \, 
\Omega^-$ candidates with $ct > 100 \, \mu$m. The projections of the 
unbinned mass fit are indicated by the dashed histograms.}
\label{fig:bbaryon_mass}
\end{figure}

\begin{figure}[h]
\centering
\includegraphics[width=80mm]{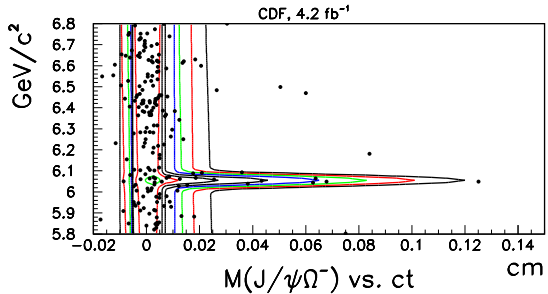}
\caption{Sigma contours in $-2\ln{\cal L}$ of the $J/\psi \, \Omega^-$ 
mass and decay time simultaneous fit.}
\label{fig:omb_sig_contour}
\end{figure}

The $J/\psi \, \Omega^-$ mass distribution with $ct \, > \,100 \, \mu$m
is shown in Fig.~\ref{fig:bbaryon_mass}(b).
The decay time ($ct$) requirement is imposed on all candidates in the
mass measurements to reduce the prompt background to the $b$-hadrons.
The significance of the 
structure seen in the $J/\psi \, \Omega^-$ mass distribution is evaluated 
with a simultaneous fit to the mass and lifetime distributions which
is maximized for two different conditions. The first maximization allows 
all parameters to vary in the fit. The second one fixes the signal 
fraction to 0.0. The value of $-2\ln{\cal L}$ obtained for the null 
hypothesis is higher than the value obtained for the fully varying 
fit by 37.3 units. We interpret this as equivalent to a $\chi^2$ 
with three degrees of freedom, which has a probability of occurrence of
$4.0\times10^{-8}$, or a $5.5\sigma$ fluctuation. 
Fig.~\ref{fig:omb_sig_contour} shows sigma contours in $-2\ln{\cal L}$ of 
the $J/\psi \, \Omega^-$ mass and decay time simultaneous fit.
We interpret the  $J/\psi \, \Omega^-$ mass distributions shown in 
Fig.~\ref{fig:bbaryon_mass}(b) to be the observation of a weakly 
decaying resonance, with a width consistent with the detector resolution. 
We treat this resonance as observation of the  $\Omega^-_b$ baryon 
through the decay process $\Omega^-_b \rightarrow J/\psi \, \Omega^-$.

\section{$\Xi^-_b$ and $\Omega^-_b$ Property Measurements
\label{sect:Properties}}
\begin{figure}[h]
\centering
\includegraphics[width=80mm]{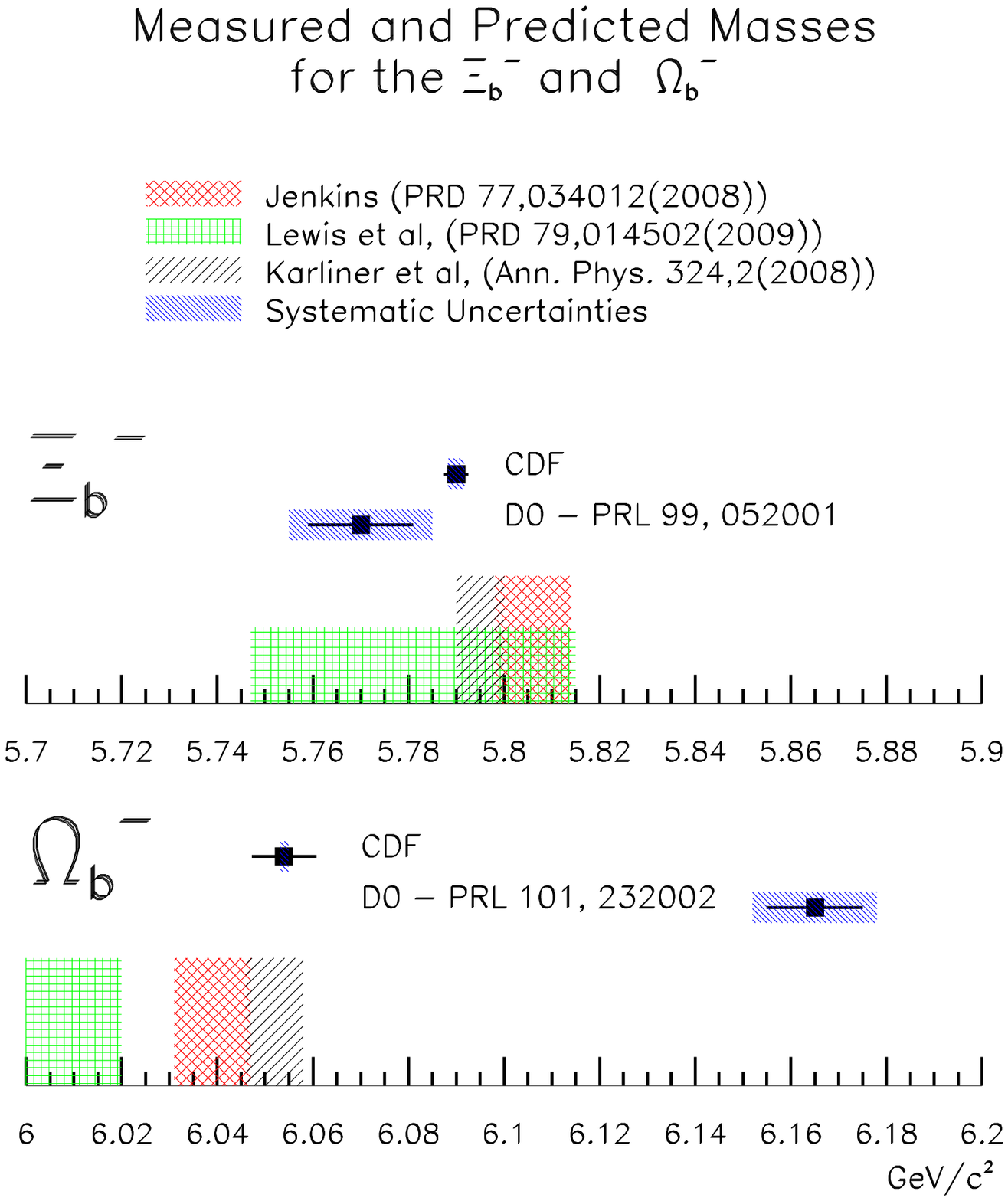}
\caption{A comparison between the mass measurements and theoretical 
predictions.}
\label{fig:mass_comparison}
\end{figure}

The mass distributions of the $\Xi^-_b$ and $\Omega^-_b$ candidates are 
shown in Fig.~\ref{fig:bbaryon_mass}, along with fit projections. The 
results of these fits as well as those from the 3 reference samples 
are listed in Table~\ref{table:properties}.
\begin{table*}[t]
\begin{center}
\caption{Measured $b$-hadrons properties.}
\begin{tabular}{|c|c|c|c|c|}
\hline
\textbf{Resonance}  & \textbf{Candidates} & \textbf{Mass  (MeV/$c^2$)} & 
\textbf{$c\tau$ ($\mu$m)} & 
\textbf{$\frac{\sigma{\cal B}}
              {\sigma(\Lambda^0_b){\cal B}(\Lambda^0_b \rightarrow
               J/\psi \, \Lambda)}$} \\
\hline
 $B^0 (J/\psi \, K^{*}(892)^{0})$ & $17520\pm305$ &  $5279.2\pm0.2$ &
$453\pm6$ & - \\
\hline
 $B^0 (J/\psi \, K^0_s) $ &  $9424\pm167$ &  $5280.2\pm0.2$ &
$448\pm7$ & - \\
\hline
 $\Lambda^0_b$              &  $1934\pm93$  &  $5620.3\pm0.5$ &
$472\pm17$ & - \\
\hline
 $\Xi^-_b $               &    $66^{+14}_{-9}$  &  $5790.9\pm2.6\pm0.8$ &
$468^{+82}_{-74}\pm0.06$ & $0.167^{+0.037}_{-0.025}\pm0.012$ \\
\hline
 $\Omega^-_b$             &    $16^{+6}_{-4}$   &  $6054.4\pm6.8\pm0.9$ &
$340^{+160}_{-120}\pm0.04$ & $0.045^{+0.017}_{-0.012}\pm0.004$\\
\hline
\end{tabular}
\label{table:properties}
\end{center}
\end{table*}
Systematic uncertainties for the $\Xi_b^-$ and $\Omega_b^-$ masses
are largely driven by our $B^0$ mass measurements, and are estimated to
be 0.8 and 0.9 MeV/$c^2$, respectively. Fig.~\ref{fig:mass_comparison}
shows our measurement of the $\Xi_b^-$ and $\Omega_b^-$ masses along with
those from the \Dzro Collaboration and the theoretical 
predictions~\cite{Theory}.
Our $\Omega_b^-$ mass result is consistent with the theoretical 
predictions and in disagreement with the \Dzro 
measurement~\cite{D0_Omega_b}.

\begin{figure}[h]
\centering
\includegraphics[width=80mm]{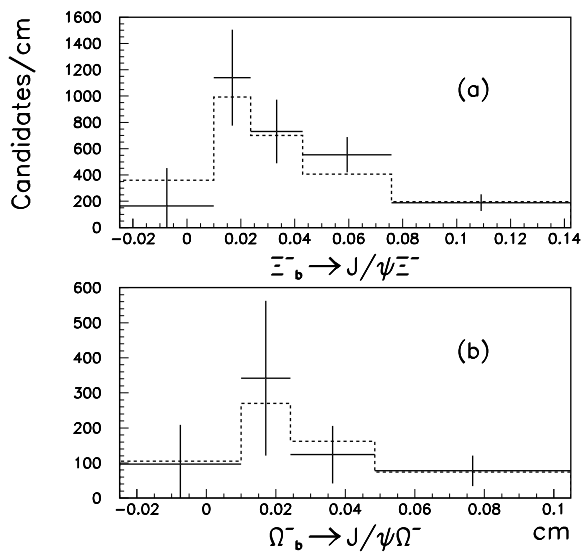}
\caption{The solid histograms represent the number of (a) $\Xi^-_b 
\rightarrow J/\psi \, \Xi^-$ and (b) $\Omega^-_b \rightarrow J/\psi \, 
\Omega^-$ candidates found in each $ct$ bin.  The dashed histogram is the
fit value.}
\label{fig:bbaryon_life}
\end{figure}

We measure $b$-hadron lifetimes by a technique which is insensitive to 
the detailed lifetime characteristics of the background. This allows 
for the lifetime calculation to be done on a relatively small sample, 
since a large number of events is not needed for background modeling.
The data are binned in $ct$, and the number of signal candidates in each
$ct$ bin is compared to the value that is expected for a particle with
a given lifetime and measurement resolution. Fig.~\ref{fig:bbaryon_life}
shows $\Xi_b^-$ and $\Omega_b^-$ candidates in $ct$ bins (solid histograms)
and their fit value (dashed histograms).
The estimates of the systematic uncertainties are obtained from the 
$B^0$ lifetime measurements. The results of the fits for the lifetimes of 
the baryons and reference samples are listed in Table~\ref{table:properties}.

Finally we present the measurements of the $\Xi_b^-$ and $\Omega_b^-$ 
production rates relative to the plentiful $\Lambda^0_b$, where we measure
ratios of cross section times branching fractions. The acceptances 
and efficiencies of the three baryon states are obtained as a function of 
$p_T$ from the detector simulation. We use the observed  $p_T$  distribution 
of $\Lambda^0_b$ production to obtain the total efficiency for the 
$\Xi^-_b$ and $\Omega^-_b$ states. The yields of the baryons extracted
from the lifetime fits are listed in Table~\ref{table:properties}, 
along with our measurements of the $\Xi^-_b$ and $\Omega_b^-$ relative 
production rates. The total relative systematic uncertainty on the production 
ratios and 7\% for the $\Xi^-_b$ and 9\% for the $\Omega_b^-$.

\section{Conclusions \label{sect:Conclusions}}
Using a 4.2 $fb^{-1}$ data sample collected with the CDF II detector at the
Tevatron we have observed a signal of $16^{+6}_{-4}$ $\Omega^-_b$ candidates,
with a significance equivalent to $5.5\sigma$. The mass, lifetime and relative 
production rates of the $\Omega^-_b$ and $\Xi^-_b$ are measured with the best 
level of precision to date. Three additional samples of $B^0$ and $\Lambda_0$ 
have been used as reference samples for cross-checks and to motivate the
systematics estimation. The masses of these baryons are in good agreement 
with the theoretical predictions, while the $\Omega^-_b$ mass is at odds with 
the previously reported measurement. More measurements are necessary to 
resolve this observed discrepancy.

\begin{acknowledgments}
We thank the Fermilab staff and the technical staffs of the participating
 institutions for their vital contributions. This work was supported by
the U.S. Department of Energy and National Science Foundation; the Italian
 Istituto Nazionale di Fisica Nucleare; the Ministry of Education, Culture,
 Sports, Science and Technology of Japan; the Natural Sciences and Engineering
 Research Council of Canada; the National Science Council of the Republic
of China; the Swiss National Science Foundation; the A.P. Sloan Foundation;
 the Bundesministerium f\"ur Bildung und Forschung, Germany; the Korean
Science and Engineering Foundation and the Korean Research Foundation;
the Science and Technology Facilities Council and the Royal Society, UK;
the Institut National de Physique Nucleaire et Physique des Particules/CNRS;
the Russian Foundation for Basic Research; the Ministerio de Ciencia e
 Innovaci\'{o}n, and Programa Consolider-Ingenio 2010, Spain;
the Slovak R\&D Agency; and the Academy of Finland.
\end{acknowledgments}

\bigskip 

\begin{thebibliography}{99} 
\bibitem{PDG} C. Amsler {\em et al.} (Particle Data Group),
  Phys. Lett. {\bf B 667}, 1 (2008).
\bibitem{D0_Xi_b} V.M. Abazov {\em et al.} (D0 Collaboration),
  Phys. Rev. Lett. {\bf 99}, 052001 (2007).
\bibitem{CDF_Xi_b} T. Aaltonen {\em et al.} (CDF Collaboration),
  Phys. Rev. Lett. {\bf 99}, 052002 (2007).
\bibitem{CDF_Sigma_b} T. Aaltonen {\em et al.} (CDF Collaboration),
  Phys. Rev. Lett. {\bf 99}, 202001 (2007).
\bibitem{D0_Omega_b} V. M. Abazov {\em et. al.} (D0 Collaboration),
  Phys. Rev. Lett. {\bf 101}, 232002 (2008).
\bibitem{CDF_Bmass}  D. Acosta {\em et al.} (CDF Collaboration),
  Phys. Rev. Lett. {\bf 96}, 202001 (2006).
\bibitem{Theory} E. Jenkins, Phys. Rev D {\bf 77}, 034012 (2008);
  R.~Lewis and R.~M.~Woloshyn, {\em ibid.} {\bf 79}, 014502 (2009);
  D.~Ebert, R.~N.~Faustov and V.~O.~Galkin, {\em ibid.} {\bf 72}, 034026 (2005);
  M. Karliner, B. Keren-Zur, H. J. Lipkin,
  and J. L. Rosner, Ann. Phys. (N.Y.) {\bf 324}, 2 (2009).
\end{thebibliography}

\end{document}